# Model-based Reinforcement Learning for Service Mesh Fault Resiliency in a Web Application-level

Fanfei Meng*
fanfeimeng2023@u.northwestern.edu
Northwestern University
USA

Lalita Jagadeesan
lalita.jagadeesan@nokia-bell-labs.com
Nokia Bell Labs
USA

Marina Thottan†
mthottan@amazon.com
Amazon Web Services
USA

## ABSTRACT
Microservice-based architectures enable different aspects of web applications to be created and updated independently, even after deployment. Associated technologies such as service mesh provide application-level fault resilience through attribute configurations that govern the behavior of request - response service – and the interactions among them – in the presence of failures. While this provides tremendous flexibility, the configured values of these attributes – and the relationships among them – can significantly affect the performance and fault resilience of the overall application. Furthermore, it is impossible to determine the best and worst combinations of attribute values with respect to fault resiliency via testing, due to the complexities of the underlying distributed system and the many possible attribute value combinations. In this paper, we present a model-based reinforcement learning workflow towards service mesh fault resiliency. Our approach enables the prediction of the most significant fault resilience behaviors at a web application-level, scratching from single service to aggregated multi-service management with efficient agent collaborations.

## 1 INTRODUCTION

A key trend in web application development in recent years is the advent of microservices-based architectures, in which applications are composed of small microservices that communicate with one another via distributed system mechanisms. Using open-source microservices technologies such as Kubernetes [1, 34, 48], developers can create and update different aspects of an application independently, even after deployment. At the same time, to ensure that faults in individual microservices – or delays in communication among them – do not cascade into application-level failures, microservice-based architectures increasingly include service mesh technologies such as Istio [2, 13] and Linkerd [3, 32]. These service meshes [12, 26, 36] contain associated "sidecars" [4] that monitor individual microservices for failures and delays, and perform associated actions to ensure application-level fault resilience; for example, bypassing problematic microservices upon consecutive errors, or ejecting them for a period of time [29, 37, 49]. The number of consecutive errors or the length of the ejection time is configured through attributes in the service mesh. The service mesh architecture is depicted in Figure 1.

As Figure 2 shows [5], multiple services are aggregated and responsible for request - connection independently, which is challenging to be optimized when faults injected [19]. Istio httpbin supported by service mesh is developed to control over web application behaviors, providing tremendous flexibility for communications. Nevertheless, the degree to which an web application is fault resilient is heavily dependent on the configured attribute values [18, 31, 47] and the relationships among them. Furthermore, it is impossible to determine the best and worst combinations of attributes via testing, due to the complexities of the underlying distributed system and the many possible attribute value combinations. The emerging machines learning methods have been widely applied to many networking problems, however, there is no previous machine learning methods focused on exploring the service mesh resiliency.

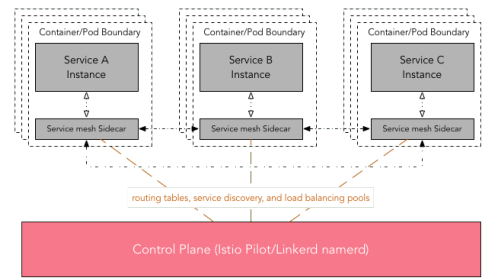

Figure 1: Service mesh architecture - CITE sidecar

In this paper, we present a model-based reinforcement learning approach towards service mesh (httpbin) fault resiliency that we call SFR2L (Service Fault Resiliency with Reinforcement Learning). Our novel contributions are as followings:

1) To the best of our knowledge, it is the first investigation on service mesh resiliency using machine learning methods, especially the first one to optimize aggregated services using model-based reinforcement learning, which enables the prediction of the most significant fault resiliency application-level behaviors.

---
*Work conducted while the author was at Nokia Bell Labs.
†Work conducted while the author was at Nokia Bell Labs.

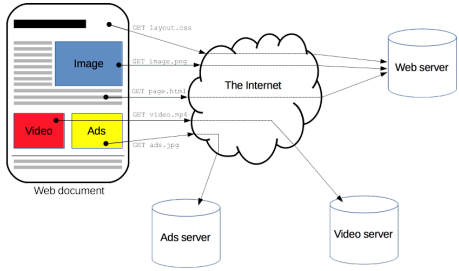

**Figure 2: Communication (request & response) between web document clients and servers, the figure only shows the request process.**

2) We develop a complete workflow to implement the flexible control over service mesh resiliency, including data collection, service modelling, policy learning for resiliency optimization and validation on simulation.
3) In terms of the policy learning, three cases are investigated: single agent for single service, multiple agents for single service and collaborative agents for multi-services, which consider most common optimizing scenarios in web applications.
4) There are **five** structured fault injection (circuit breaking pattern) datasets disclosed by us, covering from 50 to 2000 loading call settings. Each dataset is well-labelled and available at **here**. We hope this work can make contributions to the fault resiliency community and help more scholars perform more impactful researches in this field.

## 2 RELATED WORK
### 2.1 Service Modelling

[50] present a model-based reinforcement learning approach for resource allocation in scientific workflow systems based on microservices. While this paper does not address service meshes or fault resilience, our overall approach is inspired by their work.

From a service resiliency perspective, two kinds of approaches have been proposed: systematic testing and formal modeling. [24] presents an infrastructure and approach for systematic testing of resiliency. However, this work does not cover the selection of the tests to be run. Our work focuses on automating such a selection through machine learning.

In terms of formal modeling, [41] presents the use of continuous-time Markov Chains (CTMCs) and formal verification to analyze tradeoffs in service resiliency mechanisms in simple client-service interactions. Earlier work [27] also uses formal verification based on CTMCs, and analyzes multiple concurrent target services as well as steady-state availability measures including degraded service functionality.

Microservices and their self-adaptation is an active area of research, and a comprehensive survey and taxonomy of recent work is given in [20]. However, as described in [20], there is a very limited work on application-level resiliency, as well as very little work on using machine learning in the context of service self-adaptation.

**Table 1: Representative Fault Resiliency Attributes**

| Traffic & Loading Setting | Explanation |
| --- | --- |
| Max Pending Requests | Max queries on requesting connection |
| Max Connections | Max existing connection |
| Max Requests Per Connection | Max allowed requests for each connection |
| Ejection Time | The service ejected duration |
| Max Ejection | Max ejected service |
| Interval | Time between ejection and recovery |
| Consecutive Errors | Max consecutive failed requests |
| Total Threads | Max available threads |
| Total Calls | Total number of pending requests |

### 2.2 The Istio Service Mesh

As described in Section 1, Istio is an open-source service mesh technology for distributed and microservice architectures. Istio provides a transparent way to build applications. Istio's traffic management features enable service monitoring and application-level fault resilience.

In particular, Istio provides outlier detection and circuit breakers to realize fault resilience. Outlier detection enables the capacity of services to be limited when they are behaving anomalously, or even to be ejected for a period of time. Circuit breaking [6] is a capability that prevents service failures from cascading. In particular, if a service A calls another service B, which does not respond within an acceptable time period, the call can be retried or even bypassed via the circuit breaker specification. As depicted in Figure 1, the sidecar proxy for service B monitors its response, and the sidecar for service A performs the specified circuit breaking.

Istio also provides fault injection [7] and load testing [8] capabilities in order to test application fault recovery. Such testing is considered critical to perform prior to application deployment to gain confidence in the fault resilience of deployed applications.

To realize these fault resilience mechanisms, Istio enables traffic rules to be configured for application deployment; these configurations govern the specific behaviors of outlier detection and circuit breaking. Some of the attributes of these traffic rules are depicted in Table 1, and govern the number of requests and connections allowed to a service that may be behaving anomalously, the amount of time it may be ejected and at what rate and at what detection interval, and the number of consecutive errors after which a circuit breaker will be tripped. The total threads is the number of Istio worker threads, while the total calls is the number of requests to the application used in Istio load testing. Details of these attributes as well as other Istio traffic rule attributes are at [9]; configurations for fault injection and load testing are at [7, 8].

While configuration of these traffic rule attributes enables fine-grained control of fault resilience policies and fault injection and load testing, the degree to which an application is fault resilient is heavily dependent on the attribute configurations and the relationships among them. Thus, a key challenge is to determine the most significant combination of attribute values, where the "best" values from a fault resilience perspective can be used for application deployment and the "worst" values can be used to drive fault injection and load testing.

However, the determination of these most significant value combinations is highly complex due to the inter-dependencies among attributes, the failure behavior of the underlying services and the communication among them, as well as the complexities of the underlying distributed system. Thus, it is impossible to determine the most significant attribute values via manual or automated testing due to sheer number of possible behaviors.

## 3 OVERVIEW OF REINFORCEMENT LEARNING

In our context, rewards correspond to the degree of service mesh-based application over time. "Worst-case" rewards (penalties) can be used to determine configuration settings that are critical to test for resiliency in load testing prior to application deployment. For model-free methods to be used in our context, the agents must take real-time actions directly on the service mesh, and learn from the observed behavior. This necessitates implementation of algorithmic APIs in the Istio environment, likely to be very expensive and inefficient. On the other hand, model-based methods must be approached with care, as deterministic models are unlikely to yield good approximations of the transition states among different microservices.

[10] firstly propose to utilize deep neural network to generate Q-factor instead of storing large amount of reward-action pairs in harsh table. Following their works, [38] present an algorithm according to the deterministic policy gradient to execute over continuous action spaces. For model-based reinforcement learning, [23, 30, 39, 43] demonstrate the theoretical basis of policy gradient for model-based interaction [15, 16, 46]. With regard to multi-agent reinforcement learning (MARL), [44] introduce the efficient MARL algorithm for parallel policy optimization. [14] propose to deploy multi-agent to optimize the traffic controls and networks, which is an important application in actual networking practice. Communication/Collaboration is a common configuration in multi-agents system [17, 21, 28, 33, 53] and advantageous at executing more stable, efficient and better decision-makings [25, 35, 40, 45, 51, 52] using decentralized Q-network. Nevertheless, decentralized multi-agents have weak performance in the case that only small datasets are available or there are fewer state vector features for policy learning. Regarding this, [22, 27] conduct the proof of cooperative multi-agent and illustrate that the model parameters can be shared by decentralized multiple agents and agent itself also preserve its own private network to make decisions, which is the groundwork for our multi-service management. Our simulation model is inspired by [50], which introduces good approximation capabilities in assisting policy learning for service-mesh based architecture, obtaining greater training and networking optimizing efficiency than the model-free approaches.

## 4 WORKFLOW IMPLEMENTATION

Our implementation for exploring service mesh-based application fault resiliency is organized as followings (Figure 4 depicts the whole pipeline)

1) Data collection from an Istio application;
2) Simulation model training and selection;
3) Model-based Reinforcement learning;
4) Validation on policy learning results.

### 4.1 Data collection

To this end, we collect substantial traces covering target parametric spaces of the Istio httpbin service for each dataset, varying the configuration settings of the traffic rules and fault injection and load testing settings from Table 1. All collected data points are based on the actual Istio application and well-labelled after data cleaning.

### 4.2 Simulation Model

In general, we apply multiple-layer perceptrons (MLP) to simulate the aggregated behaviors, enabling agents to interact with the environment and learn the best loading space given the traffic rule attributes. Figure 3 depicts the input-output relations for modelling application-level fault resiliency. The 9-dimensional input vector contains 7 deterministic traffic rules and 2 loading settings given in Table 1, and the output corresponds to the 2 learned response attributes. We train the simulation MLP model weights: $\mathbf{W}_{SM} = \prod_{l=1}^{l}(\Phi_l + \mathbf{b}_l)$, which is the well-trained $l$-layer MLP ($\Phi_l$ is the weights of $l$-th layer and $\mathbf{b}_l$ is the corresponding bias). When the inference error reaches minimum, we save the model weights for following agent-environment interactions.

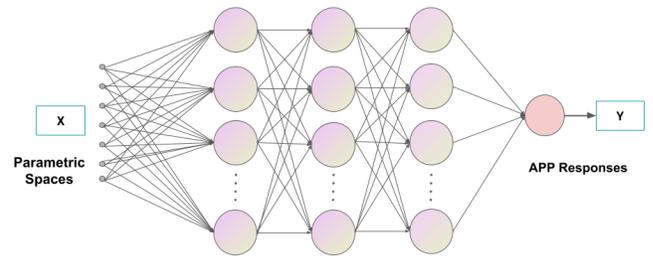

Figure 3: Simulation Model of web application-level fault resiliency. X is a 9-dimensional vector with 7 deterministic traffic rules and 2 loading settings to be decided by reinforcement learning agent(s), Y is the web application (APP) response vector with 2 features: QPS and failure rate.

### 4.3 Model-based Reinforcement Learning

We then use our simulation model as the basis for our model-based reinforcement learning (RL). In our context, the environment is our simulation model (well-trained MLP), states correspond to traffic rule settings, actions determine the loading settings (the number of threads and loading calls). We explore both single and multiple agents, the latter either working independently or dependently. As depicted in Figure 4, our agents learn from responses of our simulation model of the Istio httpbin service to the perform actions.

In real-world web application, communications between servers and clients are normally aggregated with multiple services (microservices) and holistic resiliency for all services are essential. Take a simple shopping web application for example, it may contain check-out, product, sign-up/in/out, shopping cart and review service. These services experience very closed visiting volume due to the fact that visitors would go through all these steps to finish their shopping. All services play individual roles but also have interdependencies to ensure the function of whole application. So we extend our multi-agent collaboration into management of multi-service resiliency. Section 5 describes our model-based reinforcement learning algorithm in much more detail.

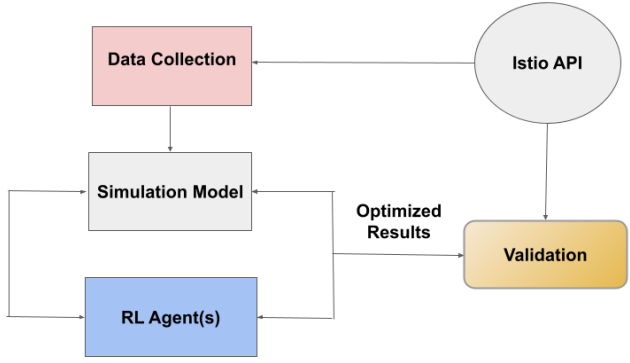

**Figure 4: SFR2L Pipeline. Data collection and validation are based on actual Istio API. RL Agent(s) fully interact with simulation model and validate its decisions in actual Istio API.**

### 4.4 Validation

Finally, we train the reinforcement learning algorithms to do policy learning and obtain optimized loading decisions given different traffic rule and loading setting combinations. All decisions with the highest reward are validated in actual Istio to evaluate policy learning.

## 5 REINFORCEMENT LEARNING PARADIGM

In the beginning, we present some preliminaries on RL and how states, actions, and rewards are defined in the context of simulation model. Based on the basic setting, we illustrate how to apply single agent to single service. Then we discuss the deployment of multi-agent reinforcement learning to address the complex parametric space optimization according to their collaborative relationships. Finally, the multi-service holistic resiliency optimization is illustrated using the combination of decentralized learning and centralized learning paradigm.

We define $t$ as the index of timestamp that RL agent interacts with the simulation model (SM), $s(t)$ as the state vector for RL agent(s) at $t$, $r(t)$ as the reward for $t$-th action $a(t)$, $\mathbf{i(t)}$ as the input vector for SM, $\mathbf{o(t)}$ as the application responses emulated by the SM. $\mathbf{W}_{SM}$ is the well-trained simulation model weights. $a(t)$ and $s(t)$ are concatenated as SM input vector $\mathbf{i}(t)$ to trigger $\mathbf{o(t)}$

$$\mathbf{i(t)} = \{\mathbf{s(t)}; a(t)\}. \quad (1)$$

; is to vertically concatenate vectors. For $\mathbf{o(t)}$

$$\mathbf{o(t)} = \mathbf{W}_{SM}\mathbf{i(t)}. \quad (2)$$

We would like to explore the borderline of normal operation of services regarding fault injection, which means higher failure rate

**Table 2: Reward Factors in Application-level**

| Reward Factors | Definition |
| --- | --- |
| Querys Per Second | Request processing speed |
| 200 Response Rate | Successful connected request rate |
| 503 Response Rate | Failed connected request rate |

(as Table 2 shows) $P_{503}(t)$ but not the full failure rate is preferred. At the same time, higher QPS ($q(t)$) is beneficial to improve the networking efficiency. So the definition of 503 reward is

$$r_{503}(t) = q(t) \cdot P_{503}(t). \quad (3)$$

### 5.1 Single Agent for Single Service

Firstly, we demonstrate the simplest case that only one agent and one simulation model interact with each other. In this case, only one kind of action (number of threads or calls) is decided by $Ag(t)$. Given a preset traffic rule $\mathbf{s(t)}$, agent $Ag(t)$ takes it as input state and make actions $a(t)$. After that, $\mathbf{i(t)}$ integrates both input (traffic rules + one loading testing) and output (optimized loading testing) of $Ag(t)$ and triggers application responses (QPS + failure rate) $\mathbf{o(t)}$. Finally, $r(t)$ is formulated from $\mathbf{o(t)}$ and update the Q-network of $Ag(t)$.

The policy of the single agent is $\pi_{\theta(t)} = a(t)$, and the Q-factor is $Q(\mathbf{s}, a) = E[r(t+1), r(t+2), ... | S(t) = \mathbf{s}, A_t = a]$, RL model is $m(\mathbf{s}, a) = E[S(t+1)|S(t) = \mathbf{s}, A_t = a]$. Our goal is to maximize the performance function $J(\theta) = E[r(1) + \alpha r(2) + \alpha^2 r(3) + ... | \pi(\theta)]$, where $\alpha$ is discounted coefficient. In the implementation, the agent will search through all the actions for a given state and select the state-action pair with the highest corresponding Q-factor using Bellman Equation [42]. The update of Q-network depends on the square error between the Q-factor that estimated by agents and actual rewards received from SM. The policy gradient for long term

$$\nabla J(\theta) = E_\gamma [\nabla_\theta \sum_{t=0}^{t-1} \log \pi(a(t)|\mathbf{s}(t)) R(\gamma)], \quad (4)$$

where $\gamma$ is the sequential action-state pair trace in time order: $\{\mathbf{s}(0), a(0), \mathbf{s}(1), a(1), ..., \mathbf{s}(t-1), a(t-1)\}$, $R(\gamma)$ is the reward function across the trace and $\nabla J(\theta)$ is the gradient used for network update. The single agent for single service is summarized in Algorithm 1.

---

**Algorithm 1:** Single Agent for Single Service

**Input:** $\mathbf{s}(1), ..., \mathbf{s}(t)$ from Round 1 to Round $t$

1 **for** $\mathbf{s}(t)$ **do**
2     Execute the $\pi_{\theta(t)}$ to clarify the optimal action $a(t)$ using $\mathbf{s}(t)$ ;
3     Combine $a(t)$ and $\mathbf{s}(t)$ to formulate input vector $\mathbf{i}(t)$ as Equation (1);
4     Obtain the SM response $\mathbf{o}_t$ as Equation (2);
5     Observe reward $r(t)$ to do policy gradient as Equation (4);
6     Proceed to next round: $t = t + 1$.

## 5.2 Multi-agents for Single Service

Section 5.1 illustrates the basic case for policy learning. However, there are two decided loading settings to tune the resiliency together. Following the previous settings, we extend the scenario into multi-agent interactions and define two kinds of collaborative relationships between two agents, which are responsible for two loading decisions, respectively. Denote $a_1(t)$ as the action taken by $Ag_1(t)$, $a_2(t)$ as the action taken by $Ag_2(t)$. $\mathbf{s}_1(t)$, $\mathbf{s}_2(t)$ are respective state vectors, $\mathbf{i}(t)$, $\mathbf{o}(t)$, $r(t)$ maintain the same definition as before. Two agents share the same reward for $\nabla J(\theta_1), \nabla J(\theta_2), \theta_1, \theta_2$ are RL Q-network parameters.

*5.2.1 Independent Collaboration.* The definition on "Independent" (Int.) originates from state vectors for two agents. In this scenario, $Ag_1(t)$ and $Ag_2(t)$ take the same state vector $\mathbf{s}(t)$ with traffic rules only and make actions in parallel: $\pi_{\theta_1}(t) = a_1(t)$ and $\pi_{\theta_2}(t) = a_2(t)$. After both actions are made, the $\mathbf{i}(t)$ is generated as

$$\mathbf{i}(t) = \{\mathbf{s}(t); a_1(t); a_2(t)\}. \tag{5}$$

After obtaining $\mathbf{i}(t)$, the following interaction with SM aligns with Section 4.1.

*5.2.2 Dependent Collaboration.* The definition on "Dependent" (Det.) also drives from the state vector for two agents. Two agents are executed in order and the state for latter agent combines the action with the state of former agent to take the second action. Assume $Ag_1(t)$ is the first executed agent, $Ag_2(t)$ is the second executed agent, state vector $\mathbf{s}_2(t)$ relies on $a_1(t)$ and $\mathbf{s}_1(t)$

$$\mathbf{s}_2(t) = \{\mathbf{s}_1(t); a_1(t)\}. \tag{6}$$

Similarly, the application response is

$$\mathbf{i(t)} = \{\mathbf{s}_2(\mathbf{t}); a_2(t)\}. \tag{7}$$

We summarize the multi-agent algorithm in Algorithm 2. $\mathbf{T}(t)$ is denoted as traffic rule vector at timestamp $t$. All types of agent interdependencies are listed in Table 3.

---

**Algorithm 2:** Multi-agent Interaction

**Input:** $\mathbf{T}(1), ..., \mathbf{T}(t)$ from Round 1 to Round $t$
1 **for** $\mathbf{T}(t)$ **do**
2    **if** *Mode = Int.* **then**
3       $a_1(t) \leftarrow \pi_{\theta_1(t)}, a_2(t) \leftarrow \pi_{\theta_2(t)}$ separately;
4       Generate $\mathbf{i(t)}$ as Equation (5);
5    **if** *Mode = Det.* **then**
6       $\mathbf{s}_1(t) = \mathbf{T}(t)$. $a_1(t) \leftarrow \pi_{\theta_1}(t)$;
7       Formulate $\mathbf{s}_2(t), a_2(t) \leftarrow \pi_{\theta_2}(t)$;
8       Generate $\mathbf{i}(t)$ as Equation (7);
9    Obtain the SM response $\mathbf{o}_t$ as Equation (2);
10   Receive reward $r(t)$ to do policy gradients for $\theta_1$ and $\theta_2$ as Equation (4);
11   Proceed to next round: $t = t + 1$.

---

**Table 3: Agents and Interdependencies. Traffic rule settings are the first 7 listed in Table 1. Thread&Call represent two agents take actions independently, Thread-Call or Call-Thread represent the actions are decided in order. The former one takes action firstly and passes it to the latter one.**

| Agent | State | Action (in order) |
|---|---|---|
| Thread Agent | 7 Traffic rules + Calls | Threads |
| Call Agent | 7 Traffic Rules + Threads | Calls |
| Thread&Call | 7 Traffic Rules for both | Threads, Calls |
| Thread-Call | 7 Rules-7 Rules + Threads | Threads, Calls |
| Call-Thread | 7 Rules-7 Rules + Calls | Calls, Threads |

## 5.3 Communicative Multi-agents for Multi-services

In light of real-world web applications subject to collaboration of all aggregated services to fulfill functions, communications are necessary to know how other services operate in the same stream. At this part, we extend our exploration that multiple services are optimized by multi-agents. However, decentralized only Q-network mandates the iterative decision execution for information sharing, which leads to serious communication latency when high volume of requests. As a consequence, we enable individual RL agents to efficiently communicate in the beginning phase so that all policy learnings can obtain better summed rewards with low time complexity.

[10, 22] propose the theoretical foundation of homogeneous agent cooperation through sharing their model parameters, allowing model to be trained with experiences of all agents, which makes the algorithm data efficient. Inspired by their work, we design that partial learning is centralized while the left learning is decentralized. The agents take different actions via searching through their own action spaces, receive rewards and update two split networks.

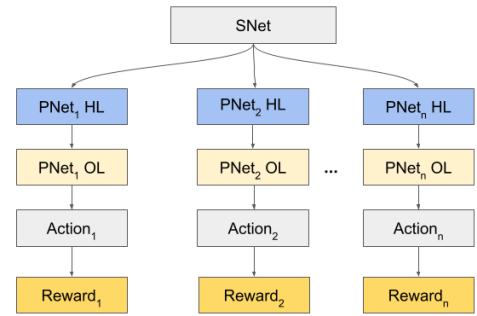

Figure 5: The configuration of Q-network of collaborative service agents: all states from all services pass the same SNet firstly, then go through individual PNet and search through their own action space to receive rewards.

We define the sharable network as SNet with input layer and its weight is $\theta_s$, decentralized network (or say private) as PNet with hidden and output layer and their weights are $\theta_{p1}, ..., \theta_{pn}$, $i$ is the index of service. As Figure 4 shows, all states from all agents pass

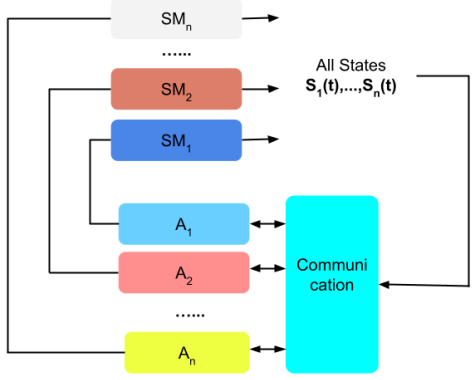

Figure 6: The interactions of multi-service using collaborative multi-agents : all same types of agents communicate with each other in the beginning, then take individual actions and finally interact with corresponding simulation models.

---

**Algorithm 3:** Collaborative Multi-services

**Input:** $T_1(1), ..., T_n(1)..., T_1(t), ..., T_n(t)$ from Round 1 to Round $t$ for $n$ services

1 **for** Round 1 to $t$: **do**
2     **for** all agents **do**
3        Generate $s_n(t)$;
4        $s_n(t)$ goes through SNet and obtain $ms_n(t)$;
5     **for** all $ms_n(t)$ **do**
6        $ms_n(t)$ goes through respective $PNet_n$.
7        $a_n(t) \leftarrow \pi_{\theta_s, \theta_{pn}(t)}$;
8        Obtain the SM response $o_n(t)$;
9        Update PNet using corresponding reward $r_n(t)$ as Equation (9);
10    Update SNet using all $r_n(t)$ as Equation (11);
11    Proceed to next round: $t = t+1$

---

SNet firstly and go through their own PNet for decision-makings. For the purpose of optimizing the holistic "worst-case" resiliency, the reward is defined as

$$r_n(t) = q_n(t) \cdot P_{503}(t) + \beta \cdot \frac{\sum_{i=0}^{n} q_n(t) \cdot P_{503}(t)}{n}, \quad (8)$$

where $\beta$ is the coefficient. After rewards are generated for each one, the respective PNet will be updated by corresponding rewards and Q-factor pair, SNet will be updated by all pairs from all service agents. As a consequence, the policy of each agent is relevant to $\theta_{pn}$ and $\pi_{\theta_s, \theta_{pn}(t)} = a_n(t)$. The long-term policy gradient for $PNet_n$

$$\nabla J(\theta_{pn}) = E_{\gamma_n} [\nabla_{\theta_{pn}} \sum_{t=0}^{t-1} \log \pi(a_n(t)|ms_n(t) R(\gamma_n)], \quad (9)$$

where $ms_n(t)$ is the output vector of SNet and the input of PNet. The prediction function $\hat{f}_{\theta_s}$ of SNet is represented by

$$s(\hat{t+1}) = \hat{f}_{\theta_s}(s_n(t), ms_n(t)). \quad (10)$$

Updating $\theta_s$ is to find the MSE minimizer of predicted $s(\hat{t+1})$ and $s(t+1)$

$$\theta_s = \underset{\theta_s}{\mathrm{argmin}} \frac{\sum_D \left\| s_n(t+1) - \hat{f}_{\theta_s}(s(t), ms_n(t)) \right\|^2}{|D|}, \quad (11)$$

where training data $s_n(t), ms_n(t), s_n(t+1) \in D$ for all agents. As shown in figure 5, the agent can firstly efficiently communicate based on shared experienced and then optimize their own decisions combining with individual case and holistic circumstances.

If both kinds of actions are decided by agents, only the same type of agents communicate over all services (call agents only communicate with other call agents, thread agent only communicate with other agents etc. The collaborative relationship for agents within one service have accordance with Section 5.2). The learning paradigm for multi-services is summarized in Algorithm 3 and visualized in Figure 6. The implementation of action space searching and PNet refreshment can be executed in parallel, only the SNet updates require iterative steps.

## 6 EXPERIMENTS
### 6.1 Simulation Model Evaluation

[11] summarizes most common ways in modelling networking communications, among which Logistic Regression, Linear Ridge Regression and Support Vector Regression are highlighted due to their outperformed simulating abilities. We select these three methods as **baseline simulation models** to fight against 5 layer-MLP and see which emulates the application response (in our case study, the Istio httpbin service) best.

We collect 5 groups of structured datasets and split them into 8:2 as training and testing set, respectively. For MLP, the input layer has 9 neurons, 3 hidden layers have 512 neurons and the output layer has 2 neurons. The learning rate varies from $1^{-6}$ to $1^{-5}$. The range for the traffic rule and thread and call settings we used in our experiments is in Table 4.

Four types of models are trained and MLP has the best performance according to Table 4, hence we use it in our approach.

### 6.2 Policy Evaluation

We explore the "worst-case" rewards, as defined in Section 5. "worst-case" rewards (penalties) provide insight into configuration settings that are critical to use in load testing prior to application deployment. The learning ability is characterized by how much RL algorithm outperform baselines under the same context, whose metrics is the maximum rolling mean ratio of cumulative reward obtained in RL to baseline in last 25 epochs.

In simulation session, we implement the simulation model to interact with RL agent(s) and record activities at the epoch with the highest reward ratio. Then data points at the best epoch work as the loading and traffic parameters to trip the actual service response and obtain the validated reward ratio. For all experiments, we analyze single agent and multiple/collaborative agent model-based reinforcement learning using our algorithm and implementation from Section 4. Every experiment is performed for 3 times and we present the mean values of 3 results.

Table 4: Ranges of Traffic Rule, Thread, and Call Settings; Model evaluation metrics: MSE (Mean Square Error).

| Traffic Rule & Loading Setting | S1 | S2 | S3 | S4 | S5 |
|---|---|---|---|---|---|
| Max Pending Requests | 1-7 | 3-7 | 12-18 | 12-18 | 15-30 |
| Max Connections | 1-7 | 3-7 | 1-5 | 10-20 | 5-15 |
| Max Req Per Connection | 1-7 | 3-7 | 10-16 | 12-18 | 15-30 |
| Ejection Time | 3m | 3m | 3m | 3m | 3m |
| Max Ejection | 100% | 100% | 4-8% | 12-18% | 22-30% |
| Interval Time | 1s | 1s | 1s | 1s | 1s |
| Consecutive Error | 1 | 1 | 4-8 | 12-18 | 22-30 |
| Total Threads | 1-5 | 3-7 | 10-16 | 12-18 | 16-20 |
| Total Calls | 400-450 | 100-700 | 50-500 | 250-600 | 1000-2000 |
| Dataset Size | 9302 | 12005 | 20592 | 12310 | 6970 |
| **Simulation Model** | **S1** | **S2** | **S3** | **S4** | **S5** |
| Support Vector Regression | 1.3 | 0.78 | 1.05 | 1.33 | 1.24 |
| Logistic Regression | 0.93 | 0.81 | 0.99 | 1.00 | 1.00 |
| Linear Ridge Regression | 0.85 | 0.84 | 0.98 | 1.01 | 0.96 |
| 5 layer-MLP | **0.13** | **0.17** | **0.52** | **0.63** | **0.38** |

Table 5: Policy Evaluations: Sim. is the maximum rolling reward ratio in simulation, Val. is the maximum rolling reward ratio in validation, *5 means 5 services are aggregated and communicative.

| Configurations | Datasets | S1 Sim. | S1 Val. | S2 Sim. | S2 Val. | S3 Sim. | S3 Val. | S4 Sim. | S4 Val. | S5 Sim. | S5 Val. |
|---|---|---|---|---|---|---|---|---|---|---|---|
| **Single for Single** | Call | 1.03 | 1.01 | 2.71 | 1.77 | 1.75 | 1.31 | 2.05 | 1.81 | 1.31 | 1.29 |
|  | Thread | 2.21 | 1.63 | 1.04 | 0.99 | 1.18 | 0.98 | 1.16 | 1.00 | 1.02 | 0.93 |
| **Multi for Single** | Thread&Call | 2.26 | 2.15 | 3.45 | 2.80 | 1.84 | 1.44 | 2.07 | 2.65 | 1.31 | 1.45 |
|  | Thread-Call | 2.24 | 2.36 | 3.39 | 2.57 | 1.96 | 1.32 | 2.14 | 3.07 | 1.32 | 1.20 |
|  | Call-Thread | 2.22 | 2.11 | 3.44 | 3.00 | 1.79 | 1.62 | 2.11 | 2.52 | 1.33 | 1.33 |
| **Multi for Multi** | Call*5 | 1.01 | 1.00 | 2.96 | 2.30 | 1.77 | 1.43 | 2.05 | 2.87 | 1.33 | 1.30 |
|  | Thread*5 | 2.23 | 1.83 | 1.15 | 1.28 | 1.13 | 1.06 | 1.18 | 1.03 | 1.01 | 1.16 |
|  | Thread&Call*5 | 2.26 | 2.35 | 4.12 | 2.92 | 1.84 | 1.29 | 2.11 | 2.83 | 1.32 | 2.05 |
|  | Thread-Call*5 | 2.22 | 1.51 | 3.53 | 2.52 | 1.94 | 2.21 | 2.09 | 3.18 | 1.34 | 1.44 |
|  | Call-Thread*5 | 2.28 | 2.19 | 3.50 | 2.33 | 1.99 | 2.40 | 2.11 | 2.16 | 1.33 | 1.44 |

We compare the selection of configuration setting values generated by our algorithm to baselines in Table 5. In terms of RL experiment, we use our deep model-based reinforcement learning algorithm to identify the configuration settings of threads and calls over time, over variations of the other traffic rule settings as per Table 1. Collaborative agent experiment includes either call or thread selection and both thread and call selection with 5 service aggregated in applications. For baseline experiment, there is no exact relevant machine learning methods addressing fault resiliency, so we randomly and evenly pick the action and see reward difference gained by non-monitoring and monitoring (RL) groups.

### 6.3 Result Analysis

From Table 5, it is clear that model-based RL algorithm is able to outperform baselines in most cases and learning effects are validated in actual Istio API. Although there are errors between simulations and validations, the ratio trends are almostly aligned. The factor of thread/call has different influence significance on the policy learning: thread is more significant for S1 and call is more significant for S2 - S5. To be extended, multi-agents learning abilities are closed to single agent when either factor is very trivial to fault resiliency, such as thread in S2, S5 and call in S1. The difference derives from the parameter range setting in simulation model training, and our approach is also beneficial to help clarify the significant factor for the service with specific traffic volume.

In terms of single service cases, most of multi-agents work better than single agent decisions, which proves that complex parameter interdependence optimization can fulfill the potential of policy learning. For instance, Thread&Call agents gain 27% higher rewards than Call only (3.45 to 2.71) agent in simulation and 69% higher rewards (2.80 to 1.77) in validation in S2. What is more important, **multi-agents usually have higher validation accuracy than single agent**. Take S1 as an example, Thread only agent has 2.21 reward ratio in simulation and 1.63 in validation (36% higher), but Thread&Call agent has closed 2.26 reward ratio and more accurate 2.15 validated ratio (5% higher). Similarly, Call only agent has 2.71 in simulation and 1.77 in validation (53% higher), Thread&Call has 3.45 in simulation and 2.80 in validation (23% higher) in S2.

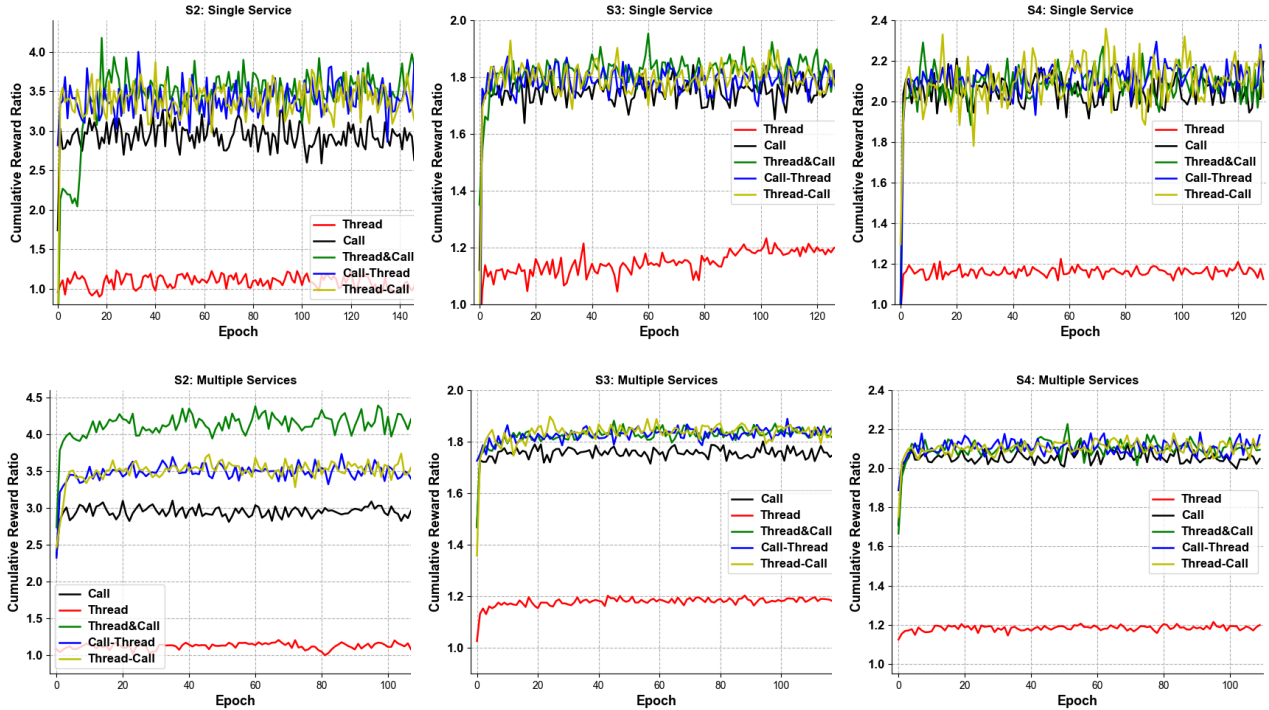

**Figure 7: Cumulative reward (per epoch) ratio. The upper are all single service case, the bottom are all aggregated multi-services. We can see that learnings are fast and then keep volatile for all cases, but multi-services with communications have more stable trends against single service. The best epoch has been reached before 150th epoch.**

Regarding the difference between single service and multi-services, it is apparent that the reward ratios obtained by all 5 services are higher than the corresponding single case and have more stable learning trends against single service, which strongly supports the importance of collaborative decision making for aggregated services resiliency, especially they experience closed loading volumes in real world. For S1, S2, multi-services outperform single-service in corresponding cases for both simulation and validation. For S3, S4 and S5, even if aggregated services behave closely to corresponding single service in simulation, the validating reward ratios are much higher. For example, we could observe that even if the reward ratios of Thread only and Call only agent in single and multiple services cases are very closed (1.13 to 1.18, 1.75 to 1.77) in S3 simulation, but S3 validating results reveal that collaborative services have better results (1.06 to 0.98, 1.43 to 1.31).

In terms of MLP approximation accuracy, Table 4 indicates that MLP has variant emulating capabilities on the behavior of service resiliency, among which S1, S2 have the lowest MSE values (0.13, 0.17). However, the difference between simulation and validation results do not strongly related to the simulating MSE value, model-based RL still provides optimized decisions on loading setting for S3 and S4 although their MLP simulating MSE values (0.52, 0.63) are relatively higher out of all groups. For model-based learning, the simulating model is not required to describe behaviors very precisely, in other words, we just need to know the estimated reactions of services given a specific context to enable continuous agent-environment interaction and policy learning. We think the experiments are also enlightening to apply model-based RL to many other fields when the real-time online interaction is expensive or unrealistic.

## 7 CONCLUSION

Our model-based reinforcement learning algorithm can help predict which values of the traffic rule settings and threads and calls yield rewards with respect to fault resiliency of the Istio httpbin service. We comprehensively investigate how model-based RL can help optimize the parametric spaces of single services, what kind of relationships between different variable agents are suitable given specific settings and how to employ service cluster using collaborative agents to communicate with each other efficiently.

In particular, the configuration settings that yield the "worst-case" rewards give insight into which combinations of configurations should be tested rigorously during load testing to ensure robust fault recovery, as these may significantly compromise application-level fault resiliency. Regarding the dynamic operating status of microservice networking, the validation on SFR2L is crucial to approve the performance of our designs. Our research is promising to the maintenance of serve mesh-based architecture in industry and construct an effective workflow to demonstrate the optimal state for a service or cluster of services.

# REFERENCES
[1] Kubernetes. https://kubernetes.io.
[2] Istio. https://istio.io.
[3] Linkerd: Homepage. https://linkerd.io.
[4] A sidecar for your service mesh. https://www.abhishek-tiwari.com/a-sidecar-for-your-service-mesh/.
[5] http in web document. https://developer.mozilla.org/en-US/docs/Web/HTTP/Overview.
[6] Circuit Breakers and Microservices Architectures. https://techblog.constantcontact.com/software-development/circuit-breakers-and-microservices/.
[7] Fault Injection. https://istio.io/latest/docs/tasks/traffic-management/fault-injection/.
[8] Circuit Breaking. https://istio.io/latest/docs/tasks/traffic-management/circuit-breaking/.
[9] Destination Rule. https://istio.io/latest/docs/reference/config/networking/destination-rule/.
[10] N. Balachandar, J. Dieter, and G. S. Ramachandran. Collaboration of AI agents via cooperative multi-agent deep reinforcement learning. *CoRR*, abs/1907.00327, 2019.
[11] R. Boutaba, M. A. Salahuddin, N. Limam, S. Ayoubi, N. Shahriar, F. E. Solano, and O. M. C. Rendon. A comprehensive survey on machine learning for networking: evolution, applications and research opportunities. *J. Internet Serv. Appl.*, 9(1):16:1–16:99, 2018.
[12] L. Calcote. *The enterprise path to service mesh architectures*. O'Reilly Media, Incorporated, 2020.
[13] L. Calcote and Z. Butcher. *Istio: Up and Running: Using a Service Mesh to Connect, Secure, Control, and Observe*. O'Reilly Media, 2019.
[14] T. Chu, J. Wang, L. Codecà, and Z. Li. Multi-agent deep reinforcement learning for large-scale traffic signal control. *IEEE Transactions on Intelligent Transportation Systems*, 21(3):1086–1095, 2020.
[15] B. B. Doll, D. A. Simon, and N. D. Daw. The ubiquity of model-based reinforcement learning. *Current opinion in neurobiology*, 22(6):1075–1081, 2012.
[16] K. Doya, K. Samejima, K.-i. Katagiri, and M. Kawato. Multiple model-based reinforcement learning. *Neural computation*, 14(6):1347–1369, 2002.
[17] T. Eccles, Y. Bachrach, G. Lever, A. Lazaridou, and T. Graepel. Biases for emergent communication in multi-agent reinforcement learning. *arXiv preprint arXiv:1912.05676*, 2019.
[18] A. El Malki and U. Zdun. Guiding architectural decision making on service mesh based microservice architectures. In *European Conference on Software Architecture*, pages 3–19. Springer, 2019.
[19] F. P. Fernandes. *Improving Web-Caching Systems with Transparent Client Support*. PhD thesis, Universidade Nova de Lisboa, 2017.
[20] M. Filho, E. Pimentel, W. Pereira, P. H. M. Maia, and M. I. Cortés. Self-adaptive microservice-based systems - landscape and research opportunities. In *2021 International Symposium on Software Engineering for Adaptive and Self-Managing Systems (SEAMS)*, pages 167–178, 2021.
[21] J. Foerster, N. Nardelli, G. Farquhar, T. Afouras, P. H. Torr, P. Kohli, and S. Whiteson. Stabilising experience replay for deep multi-agent reinforcement learning. In *International conference on machine learning*, pages 1146–1155. PMLR, 2017.
[22] J. K. Gupta, M. Egorov, and M. J. Kochenderfer. Cooperative multi-agent control using deep reinforcement learning. In *AAMAS Workshops*, 2017.
[23] J. B. Hamrick, A. L. Friesen, F. Behbahani, A. Guez, F. Viola, S. Witherspoon, T. Anthony, L. Buesing, P. Veličković, and T. Weber. On the role of planning in model-based deep reinforcement learning, 2021.
[24] V. Heorhiadi, S. Rajagopalan, H. Jamjoom, M. K. Reiter, and V. Sekar. Gremlin: Systematic resilience testing of microservices. In *2016 IEEE 36th International Conference on Distributed Computing Systems (ICDCS)*, pages 57–66, 2016.
[25] H. Hu and J. N. Foerster. Simplified action decoder for deep multi-agent reinforcement learning. *arXiv preprint arXiv:1912.02288*, 2019.
[26] K. Indrasiri and P. Siriwardena. Service mesh. In *Microservices for the Enterprise*, pages 263–292. Springer, 2018.
[27] L. J. Jagadeesan and V. B. Mendiratta. When failure is (not) an option: Reliability models for microservices architectures. In *2020 IEEE International Symposium on Software Reliability Engineering Workshops*. IEEE, 2020.
[28] J. Jiang and Z. Lu. Learning attentional communication for multi-agent cooperation. *arXiv preprint arXiv:1805.07733*, 2018.
[29] H. Johng, A. K. Kalia, J. Xiao, M. Vuković, and L. Chung. Harmonia: A continuous service monitoring framework using devops and service mesh in a complementary manner. In *International Conference on Service-Oriented Computing*, pages 151–168. Springer, 2019.
[30] L. Kaiser, M. Babaeizadeh, P. Milos, B. Osinski, R. H. Campbell, K. Czechowski, D. Erhan, C. Finn, P. Kozakowski, S. Levine, A. Mohiuddin, R. Sepassi, G. Tucker, and H. Michalewski. Model-based reinforcement learning for atari, 2020.
[31] M. Kang, J.-S. Shin, and J. Kim. Protected coordination of service mesh for container-based 3-tier service traffic. In *2019 International Conference on Information Networking (ICOIN)*, pages 427–429. IEEE, 2019.
[32] A. Khatri and V. Khatri. *Mastering Service Mesh: Enhance, secure, and observe cloud-native applications with Istio, Linkerd, and Consul.* Packt Publishing Ltd, 2020.
[33] D. Kim, S. Moon, D. Hostallero, W. J. Kang, T. Lee, K. Son, and Y. Yi. Learning to schedule communication in multi-agent reinforcement learning. *arXiv preprint arXiv:1902.01554*, 2019.
[34] M. Klein. Lyft's envoy: Experiences operating a large service mesh. 2017.
[35] S. Kumar, P. Shah, D. Hakkani-Tur, and L. Heck. Federated control with hierarchical multi-agent deep reinforcement learning. *arXiv preprint arXiv:1712.08266*, 2017.
[36] W. Li, Y. Lemieux, J. Gao, Z. Zhao, and Y. Han. Service mesh: Challenges, state of the art, and future research opportunities. In *2019 IEEE International Conference on Service-Oriented System Engineering (SOSE)*, pages 122–1225. IEEE, 2019.
[37] X. Li, X. Wang, and Y. Chen. Meshscope: a bottom-up approach for configuration inspection in service mesh. 2020.
[38] T. P. Lillicrap, J. J. Hunt, A. Pritzel, N. Heess, T. Erez, Y. Tassa, D. Silver, and D. Wierstra. Continuous control with deep reinforcement learning, 2019.
[39] Y. Luo, H. Xu, Y. Li, Y. Tian, T. Darrell, and T. Ma. Algorithmic framework for model-based deep reinforcement learning with theoretical guarantees, 2021.
[40] H. Mao, Z. Gong, Y. Ni, and Z. Xiao. Accnet: Actor-coordinator-critic net for" learning-to-communicate" with deep multi-agent reinforcement learning. *arXiv preprint arXiv:1706.03235*, 2017.
[41] N. Mendonca, C. M. Aderaldo, J. Cámara, and D. Garlan. Model-based analysis of microservice resiliency patterns. In *2020 IEEE International Conference on Software Architecture (ICSA)*, pages 114–124. IEEE, 2020.
[42] V. Mnih, K. Kavukcuoglu, D. Silver, A. Graves, I. Antonoglou, D. Wierstra, and M. Riedmiller. Playing atari with deep reinforcement learning, 2013.
[43] A. Nagabandi, G. Kahn, R. S. Fearing, and S. Levine. Neural network dynamics for model-based deep reinforcement learning with model-free fine-tuning. In *2018 IEEE International Conference on Robotics and Automation (ICRA)*, pages 7559–7566, 2018.
[44] G. Palmer, K. Tuyls, D. Bloembergen, and R. Savani. Lenient multi-agent deep reinforcement learning, 2018.
[45] P. Paquette, Y. Lu, S. Bocco, M. O. Smith, S. Ortiz-Gagné, J. K. Kummerfeld, S. Singh, J. Pineau, and A. Courville. No press diplomacy: modeling multi-agent gameplay. *arXiv preprint arXiv:1909.02128*, 2019.
[46] A. S. Polydoros and L. Nalpantidis. Survey of model-based reinforcement learning: Applications on robotics. *Journal of Intelligent & Robotic Systems*, 86(2):153–173, 2017.
[47] K. Y. Ponomarev. Attribute-based access control in service mesh. In *2019 Dynamics of Systems, Mechanisms and Machines (Dynamics)*, pages 1–4. IEEE, 2019.
[48] O. Sheikh, S. Dikaleh, D. Mistry, D. Pape, and C. Felix. Modernize digital applications with microservices management using the istio service mesh. In *Proceedings of the 28th Annual International Conference on Computer Science and Software Engineering*, pages 359–360, 2018.
[49] X. Xiaojing and S. S. Govardhan. A service mesh-based load balancing and task scheduling system for deep learning applications. In *2020 20th IEEE/ACM International Symposium on Cluster, Cloud and Internet Computing (CCGRID)*, pages 843–849. IEEE, 2020.
[50] Z. Yang, P. Nguyen, H. Jin, and K. Nahrstedt. Miras: Model-based reinforcement learning for microservice resource allocation over scientific workflows. In *2019 IEEE 39th International Conference on Distributed Computing Systems (ICDCS)*, pages 122–132, 2019.
[51] K. Zhang, Z. Yang, and T. Başar. Decentralized multi-agent reinforcement learning with networked agents: Recent advances. *arXiv preprint arXiv:1912.03821*, 2019.
[52] K. Zhang, Z. Yang, H. Liu, T. Zhang, and T. Basar. Fully decentralized multi-agent reinforcement learning with networked agents. In *International Conference on Machine Learning*, pages 5872–5881. PMLR, 2018.
[53] K. Zia, N. Javed, M. N. Sial, S. Ahmed, A. A. Pirzada, and F. Pervez. A distributed multi-agent rl-based autonomous spectrum allocation scheme in d2d enabled multi-tier hetnets. *IEEE Access*, 7:6733–6745, 2019.

Table 6: Action Spaces

| Dataset | Call | Thread |
| --- | --- | --- |
| S1 | 435, 434, ..., 450 | 1, 2, 3, 4, 5 |
| S2 | 100, 200, ..., 400 | 3, 4, 5, 6, 7 |
| S3 | 50, 100, ..., 500 | 10, 11, 12, ... , 16 |
| S4 | 250, 300, ..., 600 | 12, 13, 14, ... , 18 |
| S5 | 1000, 1100, ..., 2000 | 16, 17, 18, ... , 20 |

## 8 ADDITIONAL EXPERIMENT INFORMATION

### 8.1 Simulation Model Training

The parameter range setting for our data collection has the following criterion:

1) The failure rate and success rate distribution is relatively equivalent for whole dataset.
2) Some parameters are fixed to see better optimization on other parameters. Interval Time and Max Ejection Rate are constant in our experiments in terms of no complex topology within single httpbin service.
3) Settings are closed to cases in industrialized practice. The number of calls are set from 50 to 2000, which covers broad ranges of loading testing.

### 8.2 Training Details

We conduct our experiment for 500 epochs, in which each epoch contains 1000 simulation interactions. For the network of single agent in Section 5.1 and Section 5.2, it contains a input layer and a output layer ,and another 2 hidden layers with 512 neurons. The learning rate is $5 \times 10^{-5}$, dimension of single agent state vectors are uniformly set as 8 (7 rules + 1 loading setting), dependent multi-agent state vectors are set as 7 and 8 dimensions, independent multi-agent state vectors are set as 7 dimensions. According to Section 5.3, SNet has 2-layer 512 neurons in the end, PNet has a 512 hidden neuron in the front. The learning rate is set as $1 \times 10^{-5}$. All experiments use Adam optimizer in Pytorch library for gradient descents.

The action spaces are constrained by the datasets that used for training simulation model. The available actions are listed in Table 6.

## 9 LIMITATION

Communication latency and efficiencies in services within web applications are subject to many factors, such as the stability of internet stability, the volume of concurrent visiting and internal properties of networking (bandwidth etc.). So the validation accuracy for simulation rely on the real-time networking transmission circumstances. In the meanwhile, the data collection process is not able to precisely reflect the real-time networking operation, but the substantial traces that we collect are capable of emulating the general responses of fault injections.